\documentclass[showkeys,showpacs,prd]{revtex4}
\usepackage{amsmath}
\usepackage{amssymb}
\usepackage{graphicx}
\usepackage{color}
\usepackage{oldgerm}

\begin{document}

\title{Propagation of
gravitational waves in the nonperturbative spinor vacuum
}

\author{
Vladimir Dzhunushaliev$^{1,2,3,4}$
\footnote{v.dzhunushaliev@gmail.com}
and Vladimir Folomeev$^{4}$
\footnote{vfolomeev@mail.ru}
}
\affiliation{
$^1$ Department of Theoretical and Nuclear Physics,  Al-Farabi Kazakh National University, Almaty 050040, Kazakhstan \\
$^2$ Institute of Experimental and Theoretical Physics,  Al-Farabi Kazakh National University, Almaty 050040, Kazakhstan \\
$^3$ Institute for Basic Research, Eurasian National University, Astana, 010008, Kazakhstan \\
$^4$ Institute of Physicotechnical Problems and Material Science of the NAS
of the Kyrgyz Republic, 265 a, Chui Street, Bishkek 720071,  Kyrgyz Republic
}

\begin{abstract}
The propagation of gravitational waves on the background of a nonperturbative vacuum of a spinor field is considered.
It is shown that there are several distinctive features in comparison with the propagation of plane gravitational waves through empty space:
there exists the fixed phase difference between the
$h_{yy,zz}$ and $h_{yz}$ components of the wave;
the phase and group velocities of gravitational waves are not equal to the velocity of light;
the group velocity is always less than the velocity of light;
under some conditions the gravitational waves are either damped or absent;
for given frequency,
there exist two waves with different wave vectors.
We also discuss the possibility of experimental  verification of the obtained effects as a tool to investigate nonperurbative quantum field theories.
\end{abstract}

\date{\today}

\maketitle

\section{Introduction}

Gravitational waves (GWs) are probably the most suitable object for studying the deep space
(for a review with references on the subject see, e.g., Ref.~\cite{Centrella:2010mx}).
It is usually assumed that GWs propagate in a classical vacuum, i.e., in empty space.
But a quantum vacuum possesses the energy associated with the unavoidable quantum fluctuations of  various fields
when the vacuum expectation value of any quantum field is zero but the expectation value of the
square of fluctuations is nonzero.

In this framework, of special interest is to study the question of the propagation of GWs in the case where
fluctuations of a quantum spinor field are taken into account.
The reason is that the energy-momentum tensor of a spinor field contains
the spin connection, which in turn contains first derivatives of
tetrad components
with respect to the coordinates. As a result, the Einstein equations yield
the wave equation for a GW which contains second derivatives of the tetrad components on the
lefthand side and their first derivatives on the righthand side. 

Such a situation is a reminder of the propagation of electromagnetic waves in a continuous
conducting medium.
The corresponding wave equation is
$$
  \Delta \vec A -
  \frac{\epsilon \mu}{c^2} \frac{\partial^2 \vec A}{\partial t^2} =
  \frac{\gamma \mu}{c^2} \frac{\partial \vec A}{\partial t},
$$
where $\epsilon, \mu$ are the dielectric permittivity and magnetic permeability, respectively, $\gamma$ is the electrical conductivity.
It is well known that the above equation describes damped waves.

Comparing both these situations, one may conclude that the propagation of GWs on the background of the spinor vacuum
possesses some common features with the propagation of electromagnetic waves
in a conducting medium
(notice in this connection that the introduction of ``Ohm's gravitational law'' into the linearized Einstein equations is discussed in Ref.~\cite{Ciubotariu:1991zw}).
The problem in such studies is that
one presumably has to consider a \emph{nonperturbative} vacuum. The reason is that any perturbative calculations deal with
zero-point quantum vacuum fluctuations  of fundamental fields whose energy turns out to be infinite.
This eventually results in a number of fundamental problems, including  ultraviolet divergences and
the well-known ``cosmological constant problem'' for the Universe \cite{Sahni:1999gb}.
This motivates one to go beyond the framework of
perturbative theories
in the hope that the use of the nonperturbative quantization would allow the possibility of avoiding these problems \cite{QGQC}.

One possible way to consider a nonperturbative vacuum
might be the approach  adopted below,
which suggests a phenomenological consideration of a
nonperturbative vacuum of a spinor field.
This approach  is based on the following concepts:

(i) We make  some physically reasonable assumptions about expectation values of the spinor field and its dispersion.
Namely, we introduce an \emph{ad hoc} ans\"atz for the dispersion of the spinor field and, evaluating the covariant divergence of the
obtained righthand part of the Einstein equations, check that the Bianchi identity is satisfied. In our opinion, this
can be considered as some
approximate way to cutoff  the infinite system of differential equations for all Green's functions of the nonperturbative quantum
spinor field used in our calculations (for a more detailed discussion of this question, see Ref.~\cite{Dzhunushaliev:2013nea}).

(ii) The righthand side of the Einstein equations contains first derivatives of the tetrad with respect to the coordinates
that is a consequence of the presence of the spin connection.
As will be shown below,
the presence of the derivatives results in a fixed phase difference between the components of a GW and modification  of the dispersion relation.
Also, under some conditions the damping of GWs may arise.

Within the framework of this approach, the paper considers the simplest case of a plane GW propagating through
the nonperturbative spinor vacuum.
In this case one might consider such a vacuum as consisting of a spinor condensate (a continuous medium) through which the GW propagates.

\section{Perturbed Einstein equations}
\label{pee}

To begin with, we want to describe an exact formulation of the problem of gravitational waves propagation in a spinor vacuum.
Strictly speaking, in describing
this physical phenomenon, one needs to consider both a metric and a spinor field as quantum objects.
For the nonperturbative quantization, we have to write the following equations (for details,
see Ref.~\cite{Dzhunushaliev:2013nea})
\begin{eqnarray}
	\hat R_{\bar a \mu} - \frac{1}{2} \hat e_{\bar a \mu} \hat R
	&=& \varkappa \hat T_{\bar a \mu} ,
\label{2-0-1}\\
	\gamma^\mu \nabla_\mu \hat \psi - m \hat \psi &=& 0,
\label{2-0-2}
\end{eqnarray}
where $\hat R_{\bar a \nu}$ and $\hat R$ are correspondingly the operators  of the Ricci tensor and the Ricci scalar; $\hat e_{\bar a \mu}$ is the vierbein operator;
$\hat T_{\mu \nu}$ is the operator of the energy-momentum tensor; $\hat\psi$ is the operator of the spinor field;
$\bar a = \bar 0, \bar 1, \bar 2, \bar 3$ is the vierbein index; $\mu = 0, 1, 2, 3$ is the coordinate index;
$
\nabla_\mu \hat \psi = \partial_\mu \hat \psi -
\Gamma_\mu \hat \psi =
\partial_\mu \hat \psi +
\frac{1}{4} \hat \omega_{\bar a \bar b \mu}
\gamma^{\bar a} \gamma^{\bar b} \hat \psi
$
is the covariant derivative for the spinor with the operator of the spin connection $\hat \omega_{\bar a \bar b \mu}$~\cite{poplawski}; 
$\gamma^{\bar a}$
are the Dirac matrices in flat Minkowski spacetime; $\varkappa=8\pi \kappa/c^4$, $\kappa$ is the gravitational constant.

As of now,
a procedure of solving such an operator set of equations is unavailable. But we know that the properties of the operators are determined by all Green's functions.
For them we can write down an infinite set of equations (for details, see Ref.~\cite{Dzhunushaliev:2013nea}). Such an \emph{infinite} system of equations
 can be solved approximately by cutting it off to obtain a \emph{finite} set of equations.
Such a cut-off procedure is performed by applying some physically reasonable arguments.

Similar procedure is well known in modeling turbulence (see, for example, the textbook of Wilcox~\cite{Wilcox}). The situation there is as
follows (we follow Ref.~\cite{Wilcox} in this paragraph): One can write a statistically averaged version of the Navier-Stokes equation
(the Reynolds-averaged Navier-Stokes equation) for an averaged velocity. This equation contains six new unknown functions $\overline{\rho v_i v_j}$
(the Reynolds-stress tensor, where the overbar denotes statistical averaging). This means that our system is not yet closed. In quest of additional equations,
we have to take moments of the Navier-Stokes equation. That is, we multiply the Navier-Stokes equation by a suitable quantity and statistically
average the product. Using this procedure, we can derive a differential equation for the Reynolds-stress tensor. After such procedure we gained
six new equations, one for each independent components of the Reynolds-stress tensor. However, we have also generated 22 new unknown functions: $\overline{\rho v_i v_j v_k}$,
$
\overline{\frac{\partial u_i}{\partial x_k}
\frac{\partial u_j}{\partial x_k}
}
$,
$
\overline{
u_i \frac{\partial p}{\partial x_j}
+ u_j \frac{\partial p}{\partial x_i}
}
$.
This situation illustrates the closure problem of turbulence theory (let us note that we have the similar problem for a nonperturbative quantization).
Because of the nonlinearity of the Navier-Stokes equation, as we have higher and higher moments, we generate additional unknown functions
at each level. As written in Ref.~\cite{Wilcox}: ``The function of turbulence modeling is to derive approximations for the unknown
correlations in terms of flow properties  that are known so that a sufficient number of equations exists. In making such approximations, we close the system.''

Following this scheme, we can rephrase the last sentence as applied to a nonperturbative quantization:
\emph{The approximate  approach for a nonperturbative quantization being suggested here
is to derive approximations for unknown Green's functions using the properties of the quantum system under consideration
so that a sufficient number of equations exists. In making such approximations, we close an infinite set of equations for the Green's functions.}

Here we employ some approximation, as described below.
We evaluate the Bianchi identities,
instead of solving the Dirac equation which contains the nonlinear term
$
\left\langle
	\hat{\bar \psi} \hat \omega_{\bar a \bar b \mu} \hat \psi
\right\rangle$.
The presence of
this term prevents us from solving the operator Dirac equation directly since, as follows from the above discussion,
to do this
we have to write an equation for
the term
$
\left\langle
	\hat{\bar \psi} \hat \omega_{\bar a \bar b \mu} \hat \psi
\right\rangle
$,
and so on {\it ad infinitum}. This makes us use the Bianchi identities instead of the Dirac equation. Nevertheless, if one wants to continue with calculations
in the next  approximation, the Dirac equation will necessarily appear.

The operator of the energy-momentum tensor of the spinor field is given by
\begin{equation}
  \hat T_{\bar a \bar b} = \frac{i}{2} \left[
    \hat {\bar \psi} \gamma_{(\bar a} \nabla_{\bar b )} \hat \psi
  - \nabla_{(\bar a} \hat {\bar \psi} \gamma_{\bar b)} \hat \psi
  \right ] +
  \eta_{\bar a \bar b} \left(
  		- \frac{i}{2}
  		\hat {\bar \psi} \gamma^\mu \nabla_\mu \hat \psi +
  		\frac{i}{2} 	\nabla_\mu \hat {\bar \psi} \gamma^\mu \hat \psi
 		+ m \hat {\bar \psi} \hat \psi
  \right) ,
\label{2-0-3}
\end{equation}
where $\hat {\bar \psi} \gamma_{(\bar a} \nabla_{\bar b )}$ means the symmetrization over the indices $\bar a, \bar b$; $m$ is the mass of the spinor field.

Equations \eqref{2-0-1} and  \eqref{2-0-2} cannot be solved explicitly, and we have to use some approximation. First let us write down the expectation value of these equations
\begin{eqnarray}
	\left\langle Q \left|
		\hat R_{\bar a \mu} - \frac{1}{2} \hat e_{\bar a \mu} \hat R
	\right| Q \right\rangle
	&=& \varkappa \left\langle Q \left| \hat T_{\bar a \mu}
	\right| Q \right\rangle 		,
\label{2-0-4}\\
	\left\langle Q \left|
		\gamma^\mu \nabla_\mu \hat \psi - m \hat \psi
	\right| Q \right\rangle &=& 0,
\label{2-0-5}
\end{eqnarray}
where $\left. \left. \right| Q \right\rangle$ is a quantum state describing the propagation of a GW through a spinor vacuum.
Let us note that as the consequence of \eqref{2-0-5}, the expectation value of the term in the parentheses of Eq.~\eqref{2-0-3} is exactly zero.
Once again, we emphasize that we cannot use the Dirac equation \eqref{2-0-5} to calculate the expectation value of the spinor field
$\left\langle \hat \psi \right\rangle$, since, as mentioned
at the beginning of this section, strictly speaking, in performing such calculations
we must also quantize a metric. In this case the expectation value of the Dirac equation will contain not only
$\left\langle \hat \psi \right\rangle$ but also the term
$\left\langle \hat \omega_{ab \mu} \hat \psi \right\rangle$. Then
we will have to write down a new equation for this Green's function, and so on {\it ad infinitum}. This is the main problem  encountered in
 Heisenberg's nonperturbative quantization technique, discussed also in Ref.~\cite{Dzhunushaliev:2013nea}. To avoid this problem, we employ the aforementioned approximation.

To solve Eqs.~\eqref{2-0-4} and  \eqref{2-0-5}, we assume the following approximations: (a) the vierbein
$e^{\bar a}_{\phantom{a} \mu}$ and all geometrical quantities (the Ricci tensor, the Ricci scalar, and the spin connection) are the classical ones; (b) instead of solving
the Dirac equation \eqref{2-0-5}, we will check the validity of the Bianchi identities for the righthand side of the Einstein equations~\eqref{2-0-4} with the shortened
energy-momentum tensor; (c) we consider only weak GWs.

Within our approximation, we will consider the following set of equations:
\begin{eqnarray}
  \delta R_{\bar a \bar b} -
  \frac{1}{2} \eta_{\bar a \bar b} \delta R &=&
  \varkappa \left\langle Q \left|
    \widehat{\delta T}_{\bar a \bar b}
  \right| Q \right\rangle,
\label{2-0-10}\\
	\left\langle Q \left|
		\widehat{\delta T}_{\bar a}^{\phantom a \mu}
	\right| Q \right\rangle_{; \mu} &=& 0,
\label{2-0-12}
\end{eqnarray}
where $\delta R_{\bar a \bar b}$ and $\delta R$ are the gravitational wave approximation for the Ricci tensor and the Ricci scalar, as given below by Eq.~\eqref{1-220}.
In turn, the righthand side of Eq.~\eqref{2-0-10} is calculated
in subsequent sections.  To simplify the notation we will hereafter use $\left\langle \cdots \right\rangle$ instead of
$\left\langle Q \left| \cdots
\right| Q \right\rangle$.

\subsection{The lefthand side of the perturbed Einstein equations}

According to Ref.~\cite{AndrewHamilton}, let the vierbein perturbation $\phi_{\bar a}^{\phantom a \bar b}$ is defined in the following manner:
\begin{eqnarray}
  e^{\bar a}_{\phantom{a}\mu} = \left(
    \delta^{\bar a}_{\bar b} - \phi^{\bar a}_{\phantom a \bar b}
  \right) \stackrel{0}{e}{\!}^{\bar b}_{\phantom{a}\mu},
\label{a1-10}\\
  e_{\bar a}^{\phantom{a}\mu} = \left(
    \delta_{\bar a}^{\bar b} + \phi_{\bar a}^{\phantom a \bar b}
  \right) \stackrel{0}{e}{\!}_{\bar b}^{\phantom{b}\mu},
\label{a1-20}
\end{eqnarray}
where $\stackrel{0}{e}{\!}^{\bar a}_{\phantom{a}\mu}$ is the unperturbed tetrad; $\stackrel{0}{e}{\!}_{\bar a}^{\phantom{b}\mu}$
is the unperturbed inverse tetrad; $e^{\bar a}_{\phantom{a}\mu}$ is the perturbed tetrad;
$e_{\bar a}^{\phantom{a}\mu}$ is the perturbed inverse tetrad;
$- \phi^{\bar a}_{\phantom a \bar b} \stackrel{0}{e}{\!}^{\bar b}_{\phantom{a}\mu}$ is the perturbation of the tetrad. It is convenient
to work with the covariant tetrad-frame components
$\phi_{\bar a \bar b}$ of the tetrad perturbation
\begin{equation}
  \phi_{\bar a \bar b} = \eta_{\bar b \bar c}
  \phi_{\bar a}^{\phantom{a} \bar c},
\label{a1-30}
\end{equation}
where $\eta_{\bar a \bar b} = \mathrm{diag}\left\{ +,-,-,- \right\}$ is the Minkowski metric. For a single Fourier mode,
whose wave vector $\vec{k}$ is taken to lie in the $x$-direction we have
\begin{equation}
  \phi_{\bar a \bar b} =
  \left(
    \begin{array}{cccc}
      \psi & \partial_x w & w_{\bar 2} & w_{\bar 3} \\
      \partial_x \tilde w & \Phi + \partial_x^2 h &
      \partial_x h_{\bar 2} & \partial_x h_{\bar 3} \\
      \tilde w_{\bar 2} & \partial_x \tilde h_{\bar 2} &
      \Phi + h_{\bar 2 \bar 2} & h_{\bar 2 \bar 3} +
      \partial_x \tilde h \\
      \tilde w_{\bar 3} & \partial_x \tilde h_{\bar 3} &
      h_{\bar 3 \bar 2} - \partial_x \tilde h & \Phi -
      h_{\bar 2 \bar 2} \\
    \end{array}
  \right).
\label{a1-120}
\end{equation}
One can introduce the gauge invariant functions $\Psi$ and  $W_i$
\begin{eqnarray}
  \Psi &=& \psi - \partial_t \left(
    w + \tilde w - \partial_t h
  \right),
\label{a1-130}\\
  W_i &=& w_i + \tilde w_i - \partial_t \left(
    h_i + \tilde h_i
  \right),
\label{a1-140}
\end{eqnarray}
where $i=1,2,3$ are the spacelike world indices. After that the perturbations of the Einstein tensor are
\begin{equation}
  \delta G_{\bar a \bar b} =
  \left(
    \begin{array}{cccc}
      - 2 \partial_x^2 \Phi &
      2 \partial_x \dot \Phi &
      - \frac{1}{2} \partial^2_x W_y &
      - \frac{1}{2} \partial^2_x W_z \\
      2 \partial_x \dot \Phi & - 2 \ddot \Phi &
      \frac{1}{2} \partial_x \dot W_y &
      \frac{1}{2} \partial_x \dot W_z \\
      - \frac{1}{2} \partial^2_x W_y &
      \frac{1}{2} \partial_x \dot W_y &
      - 2 \ddot \Phi - \partial^2_x \left( \Psi - \Phi \right) + \Box h_+&
      \Box h_\times \\
      - \frac{1}{2} \partial^2_x W_z &
      \frac{1}{2} \partial_x \dot W_z &
      \Box h_\times &
      - 2 \ddot \Phi - \partial^2_x \left( \Psi - \Phi \right) - \Box h_+
      \\
    \end{array}
  \right),
\label{1-220}
\end{equation}
where the dot denotes differentiation with respect to $\tau=c t$;
$\Box = \frac{\partial^2}{\partial \tau^2} - \nabla^2$ is the d$\acute{}$ Alembertian and $h_+$ and $h_\times$ are the two polarizations of gravitational waves
\begin{equation}
  h_+ = h_{yy} = -h_{zz}, \;
  h_\times = h_{yz} = h_{zy},
\label{1-230}
\end{equation}
and $\Phi, W_{\bar i}, \Psi$ are vierbein components given by the formulae \eqref{a1-120}-\eqref{a1-140}.

\subsection{The righthand side of the Einstein equations}
\label{RHS_En}

To calculate the expectation value of the energy-momentum tensor of the spinor field, we state the following assumptions concerning the spinor field:
\begin{itemize}
  \item The vacuum expectation value of the spinor field is zero:
  \begin{equation}
    \left\langle
      \hat \psi_{\mathfrak a}
    \right\rangle = 0.
  \label{1-2-10}
  \end{equation}
  \item The vacuum expectation value of the product of the spinor field in two points $x,y$ is nonzero:
  \begin{equation}
    \left\langle
      \hat \psi_{\mathfrak a}^*(x) \hat \psi_{\mathfrak b}(y)
    \right\rangle =
    \Upsilon_{\mathfrak a \mathfrak b}(x,y) \neq 0
  \label{1-2-20}
  \end{equation}
	here $\hat \psi$ is the operator of the spinor field;
	$\mathfrak a, \mathfrak b$ are 	the spinor indices;
	$\Upsilon_{\mathfrak a \mathfrak b}$
	is the 2-point Green's function.
  \item Every component
  \begin{equation}
   \left| \Upsilon_{\mathfrak a \mathfrak b}(x,y) \right|= \mathrm{const.}
  \label{1-2-25}
  \end{equation}
  \item As a consequence of Eq.~\eqref{1-2-25} we have
  \begin{equation}
   \left\langle
      \hat \psi_{\mathfrak a}^*(x)
      \partial_{y^\mu}\hat \psi_{\mathfrak b}(y)
   \right\rangle = 0.
  \label{1-2-27}
  \end{equation}
\end{itemize}
The energy-momentum tensor contains the following unperturbed and perturbed contributions:
\begin{equation}
  \hat T_{\bar a \bar b} = \stackrel{0}{\hat T}_{\bar a \bar b} +
  \widehat{\delta T}_{\bar a \bar b},
\label{1-2-50}
\end{equation}
where $\stackrel{0}{\hat T}_{\bar a \bar b}$ is calculated for unperturbed Minkowski spacetime with zero spin connection, $\omega_{\bar a \bar b \mu} = 0$.
 Consequently,
\begin{equation}
  \stackrel{0}{\hat T}_{\bar a \bar b} = \frac{i}{2} \left[
    \hat {\bar \psi} \gamma_{(\bar a} \partial_{\bar b )} \hat \psi
  - \partial_{(\bar a} \hat {\bar \psi} \gamma_{\bar b)} \hat \psi
  \right ].
\label{1-2-60}
\end{equation}
According to Eq.~\eqref{1-2-27},
\begin{equation}
	\left\langle
  		\stackrel{0}{\hat T}_{\bar a \bar b}
	\right\rangle = 0.
\label{1-2-70}
\end{equation}
Its physical meaning is that since the expectation value of
the energy-momentum tensor in unperturbed Minkowski spacetime is equal to zero,  it does not affect the propagation of GWs.
The perturbed energy-momentum tensor is calculated in Appendix~\ref{app1}.

\section{Gravitational wave propagating on the background of the spinor vacuum}

We consider a GW propagating along the $x$ axis, described by
the Einstein equations~\eqref{2-0-10}.
It has to be emphasized that the righthand side of these equations cannot be calculated by using a perturbative technique.
The reason is that perturbative calculations give us an infinite energy of zero-point vacuum fluctuations.
This energy  acts as a source of gravitational field and, in general, cannot be excluded by using a renormalization procedure~\cite{Birrell}.
In fact, this is just an imprint of the well-known problem of the contradiction between gravity and the \emph{perturbative}
quantum paradigm.

To calculate a nonperturbative expectation value of
$\left\langle \hat T_{\bar a \bar b} \right\rangle$,
we will use the
assumptions about expectation values of the spinor field and its dispersion as described in the previous section. In doing so,
we will consider a particular case of GWs for which
\begin{equation}
  \Phi = \Psi = W_i = 0.
\label{1-3-20}
\end{equation}
Below we consider two different ans\"atzs for the spinor field.

\subsection{Case I}
\label{case1}

For the ans\"atz
\begin{equation}
 \hat\psi = e^{-i (\omega t - k x)}\begin{pmatrix}
		\hat A 	\\
		\hat B 		\\
		\hat B	\\
		\hat A
	\end{pmatrix},
\label{2a-10}
\end{equation}
where $\omega$ is the frequency and $k$ is the $x$-component of the wave vector.

The algorithm for calculating  the righthand sides of Eqs. \eqref{2-0-10} is as follows. The first step is to evaluate them as
classical quantities using \eqref{2a-10} without hats over $A,B$, and then to restore the hats: $\hat A, \hat B$.
These calculations give the $\bar a, \bar b \neq \bar y, \bar z$ components of the classical energy-momentum tensor $T_{\bar a \bar b}$ which contain only the terms
\begin{equation}
	 A^*  B -  A  B^* +  S  V^* -  S^*  V \quad
	\text{and} \quad
	\left|  A \right|^2 - \left| B \right|^2 -
	\left| S \right|^2 + \left| V \right|^2.
\label{a1-pert_eq_2a}
\end{equation}
In calculating components of the energy-momentum tensor, we have used the spinor in the general form
$\psi^{T}=e^{-i \left( \omega t - k x \right)}(A,B,V,S)$.
Taking into account the gauge~\eqref{1-3-20},
the lefthand side of the Einstein equations~\eqref{2-0-10} is not zero only for the $yy, zz, yz$ components. Therefore we have to choose
$\hat A, \hat B, \hat S, \hat V$ in such a manner that
the only nonzero components of
the energy-momentum tensor would be
$
T_{\bar y \bar y, \bar z \bar z, \bar y \bar z}$. We see  from \eqref{a1-pert_eq_2a} that the components
$
T_{\bar a \bar b},
(\bar a \bar b \neq \bar y \bar y, \bar z \bar z, \bar y \bar z)
$ of the energy-momentum tensor are equal to zero only when
\begin{eqnarray}
	 \text{Case I:}&& \quad V =  B,  S =  A; 
\label{2a-12a}\\
	 \text{Case II:}&& \quad B =  A,  S =  V.
\label{2a-12b}
\end{eqnarray}
In this subsection we consider the first case, corresponding to the ans\"atz~\eqref{2a-10},
and the second case, corresponding to the ans\"atz~\eqref{2b-10}, will be studied in next subsection.
We assume the following values of the 2-point Green's functions of the spinor field $\psi$:
\begin{eqnarray}
  \Upsilon &=& \left\langle \psi_1^* \psi_2 \right\rangle =
  \left\langle \psi_1^* \psi_3 \right\rangle =
  \left\langle \psi_4^* \psi_2 \right\rangle =
  \left\langle \psi_4^* \psi_3 \right\rangle =
  \left\langle A^* B \right\rangle = \Upsilon_1 + i \Upsilon_2,
\label{1a-3-50}\\
  \Upsilon^* &=& \left\langle \psi_2^* \psi_1 \right\rangle =
  \left\langle \psi_2^* \psi_4 \right\rangle =
  \left\langle \psi_3^* \psi_1 \right\rangle =
  \left\langle \psi_3^* \psi_4 \right\rangle =
  \left\langle B^* A \right\rangle = \Upsilon_1 - i \Upsilon_2,
\label{1a-3-60}
\end{eqnarray}
with $\left| \Upsilon_{1,2} \right| =\text{const}$. By choosing  $\hat A, \hat B, \hat S, \hat V$
in the form of \eqref{2a-12a},
the
$\bar y \bar y, \bar z \bar z,$ and $\bar y \bar z$ components of the energy-momentum tensor \eqref{1-2-80} are
\begin{eqnarray}
	 \left\langle
	 	\widehat{\delta T_{\bar y \bar y}}
	 \right\rangle &=& -
	 \left\langle
	 \widehat{\delta T_{\bar z \bar z}}
	 \right\rangle =
	 2 \left(
	 	\left\langle \hat A^* \hat B + \hat A \hat B^* \right\rangle
	 \right) \dot h_{\bar y \bar z},
\label{1a-3-62}\\	
	\left\langle
	 \widehat{\delta T_{\bar y \bar z}}
	\right\rangle &=& 2 \left(
	 	\left\langle \hat A^* \hat B + \hat A \hat B^* \right\rangle
	 \right) \dot h_{\bar y \bar y}.
\label{1a-3-64}
\end{eqnarray}
Equations~\eqref{2-0-10} with the gauge \eqref{1-3-20} and the perturbed components of the energy-momentum tensor \eqref{1a-3-62} and \eqref{1a-3-64}
give the following set of equations for the components \eqref{1-230}:
\begin{eqnarray}
 h_{\bar y \bar y}^{\prime \prime}-\ddot{h}_{\bar y \bar y} &=& -2 \varkappa
 \left(
  \left\langle \hat A^* \hat {B} \right\rangle +
  \left\langle \hat A \hat {B^*} \right\rangle
 \right) \dot h_{\bar y \bar z} ,
\label{a1-pert_eq_1}\\
 h_{\bar y \bar z}^{\prime \prime}-\ddot{h}_{\bar y \bar z} &=& 2\varkappa \left(
  \left\langle \hat A^* \hat {B} \right\rangle +
  \left\langle \hat A \hat {B^*} \right\rangle
 \right) \dot h_{\bar y \bar y},
\label{a1-pert_eq_2}
\end{eqnarray}
where the prime denotes differentiation with respect to $x$,
and the appearance of the derivatives of the components ${h}_{\bar y \bar y}, {h}_{\bar y \bar z}$ on the righthand side of these equations is
connected with the presence of the spin connection on the righthand side of Einstein's equations.

We are looking for the $x$-plane wave solution in the form
\begin{eqnarray}
  h_{\bar y \bar y} = - h_{\bar z \bar z} &=& A_1 e^{-i \left( \omega t - k x \right)},
\label{1a-3-30}\\
  h_{\bar y \bar z} = h_{\bar z \bar y} &=& A_2 e^{-i \left( \omega t - k x \right)}.
\label{1a-3-40}
\end{eqnarray}
Substituting the solutions \eqref{1a-3-30} and \eqref{1a-3-40} into
the wave equations \eqref{a1-pert_eq_1} and  \eqref{a1-pert_eq_2} and using the expressions \eqref{1a-3-50} and \eqref{1a-3-60}, we obtain the following relations
(hereafter we work in natural units where $\hbar=c=1$):
\begin{eqnarray}
  A_1 \left( k^2 - \omega^2 \right) &=& - 4 i \varkappa A_2 \Upsilon_1 \omega ,
\label{1a-3-70}\\
  A_2 \left( k^2 - \omega^2 \right) &=& 4 i \varkappa A_1 \Upsilon_1 \omega .
\label{1a-3-80}
\end{eqnarray}
From them one can immediately read out
\begin{equation}
  A_2 = \pm i A_1 = A_1 e^{\pm i \frac{\pi}{2}}.
\label{1a-3-90}
\end{equation}
This means that the phase difference between $\bar y \bar y, \bar z \bar z$ and $\bar y \bar z$ components of the GW is $\pm \pi/2$. In turn, the dispersion relation is
\begin{equation}
  k^2 = \omega^2 \pm 4 \varkappa \Upsilon_1 \omega .
\label{1a-3-100}
\end{equation}
Thus, we see that there are two GWs with different wave vectors for the same frequency $\omega$.
But for the case $k=\sqrt{\omega^2 - 4 \varkappa \Upsilon_1 \omega}$ a situation may occur where the GW does not exist. This happens if
\begin{equation}
  \omega < 4 \varkappa \Upsilon_1 .
\end{equation}
The phase velocity of the GW is given by
\begin{equation}
  v_{p} = \frac{\omega}{k} =
  \sqrt{\frac{1}{1 \pm \frac{4 \varkappa \Upsilon_1}{\omega}}}
  \neq 1
\label{1a-3-110}
\end{equation}
(recall that $v$ is measured in units of $c$). We see that there are two branches: one with $v_{p} < 1$ and the other with $v_{p} > 1$.

The group velocity of the GW is
\begin{equation}
  v_{g} = \frac{d \omega}{d k} =
  \frac{\sqrt{1 \pm \frac{4\varkappa \Upsilon_1}{\omega}}}
  {1 \pm \frac{2 \varkappa \Upsilon_1}{\omega}}
  \neq 1.
\label{1a-3-120}
\end{equation}
It is interesting that if $\frac{\varkappa \Upsilon_1}{\omega} \ll 1$ then $v_{g} \approx 1$.
It is also seen that the group velocity $v_{g}< 1$ for any value of $\varkappa \Upsilon_1/\omega$ and for any sign of $\Upsilon_1$.

\subsection{Case II}

In this section we consider the following ans\"atz for the spinor field:
\begin{equation}
 \hat\psi = e^{-i (\omega t - kx)}\begin{pmatrix}
		\hat A 	\\
		\hat A 		\\
		\hat V	\\
		\hat V
	\end{pmatrix},
\label{2b-10}
\end{equation}
where
$
\left\langle \hat A \hat {A^*} \right\rangle,
\left\langle \hat V \hat {V^*} \right\rangle
$
are taken to be constant in accordance with the assumption \eqref{1-2-25}.

For the ans\"atz \eqref{2b-10}, we assume the following values of the 2-point Green's functions of the spinor field $\psi$:
\begin{eqnarray}
  \left\langle \psi_1^* \psi_2 \right\rangle =
  \left\langle \psi_1 \psi_2^* \right\rangle =
  \left\langle \psi_1^* \psi_1 \right\rangle =
  \left\langle \psi_2^* \psi_2 \right\rangle =
  \left\langle \hat A^* \hat A \right\rangle &=&
  \Upsilon_1,
\label{2b-12}\\
  \left\langle \psi_3^* \psi_4 \right\rangle =
  \left\langle \psi_4^* \psi_3 \right\rangle =
  \left\langle \psi_3^* \psi_3 \right\rangle =
  \left\langle \psi_4^* \psi_4 \right\rangle =
  \left\langle \hat V^* \hat V \right\rangle &=&
  \Upsilon_2,
\label{2b-14}
\end{eqnarray}
with $\left| \Upsilon_{1,2} \right| = \text{const}.$ By choosing
 $\hat A, \hat B, \hat S, \hat V$ in the form of \eqref{2a-12b},
 the
$\bar y \bar y, \bar z \bar z$ and $\bar y \bar z$ components of the energy-momentum tensor \eqref{1-2-80} are
\begin{eqnarray}
	 \left\langle
	 	\widehat{\delta T_{\bar y \bar y}}
	 \right\rangle &=& -
	 \left\langle
	 \widehat{\delta T_{\bar z \bar z}}
	 \right\rangle =
	 2 \left[ \left(
  \left\langle \hat A \hat {A^*} \right\rangle -
  \left\langle \hat V \hat {V^*} \right\rangle
 \right)h_{\bar y \bar z}^{\prime} + \left(
  \left\langle \hat A \hat {A^*} \right\rangle +
  \left\langle \hat V \hat {V^*} \right\rangle
 \right) \dot{h}_{\bar y \bar z} \right],
\label{2a-11}\\	
	\left\langle
	 \widehat{\delta T_{\bar y \bar z}}
	\right\rangle &=& - 2 \left[ \left(
  \left\langle \hat A \hat {A^*} \right\rangle -
  \left\langle \hat V \hat {V^*} \right\rangle
 \right)h_{\bar y \bar y}^{\prime} +
 \left(
  \left\langle \hat A \hat {A^*} \right\rangle +
  \left\langle \hat V \hat {V^*} \right\rangle
 \right) \dot{h}_{\bar y \bar y} \right].
\label{2a-12}
\end{eqnarray}
Substituting these expressions
into Eq.~\eqref{2-0-10} and taking into account the gauge \eqref{1-3-20},
we have the following set of equations for the components \eqref{1-230}:
\begin{eqnarray}
 h_{\bar y \bar y}^{\prime \prime}-\ddot{h}_{\bar y \bar y}&=&-2\varkappa \left[
 \left(
  \left\langle \hat A \hat {A^*} \right\rangle -
  \left\langle \hat V \hat {V^*} \right\rangle
 \right)h_{\bar y \bar z}^{\prime} + \left(
  \left\langle \hat A \hat {A^*} \right\rangle +
  \left\langle \hat V \hat {V^*} \right\rangle
 \right)\dot{h}_{\bar y \bar z}
 \right],
\label{pert_eq_1}\\
 h_{\bar y \bar z}^{\prime \prime}-\ddot{h}_{\bar y \bar z}&=&2\varkappa \left[
 \left(
  \left\langle \hat A \hat {A^*} \right\rangle -
  \left\langle \hat V \hat {V^*} \right\rangle
 \right)h_{\bar y \bar y}^{\prime} +
 \left(
  \left\langle \hat A \hat {A^*} \right\rangle +
  \left\langle \hat V \hat {V^*} \right\rangle
 \right)\dot{h}_{\bar y \bar y}
 \right].
\label{pert_eq_2}
\end{eqnarray}
The algorithm for calculating the righthand sides of
these equations is the same as that for the case I from subsection~\ref{case1}.
The appearance of the derivatives of the components ${h}_{yy}, {h}_{yz}$ on the righthand side of these equations, as before,
is connected with the presence of the spin connection on the righthand side of Einstein's equations.

Again, we are looking for the $x$-plane wave solution in the form \eqref{1a-3-30} and \eqref{1a-3-40}.
Substituting them into  the wave equations \eqref{pert_eq_1} and \eqref{pert_eq_2}
and taking into account \eqref{2b-12} and \eqref{2b-14}, we obtain the following relations:
\begin{eqnarray}
  A_1 \left( k^2 - \omega^2 \right) &=& 2 i \varkappa A_2 \left[
   k \left( \Upsilon_1 - \Upsilon_2 \right) -
   \omega \left( \Upsilon_1 + \Upsilon_2 \right)
  \right] ,
\label{1-3-60}\\
  A_2 \left( k^2 - \omega^2 \right) &=& - 2 i \varkappa A_1 \left[
   k \left( \Upsilon_1 - \Upsilon_2 \right) -
   \omega \left( \Upsilon_1 + \Upsilon_2 \right)
  \right],
\label{1-3-70}
\end{eqnarray}
which immediately give
\begin{equation}
  A_2 = \pm i A_1 = A_1 e^{\pm i \frac{\pi}{2}}.
\label{1-3-80}
\end{equation}
That is, the phase difference between
$\bar y \bar y, \bar z \bar z$ and $\bar y \bar z$ components of the GW is again $\pm \pi/2$, as in the case I. In turn, the dispersion relation takes the form
\begin{equation}
  k^2 \pm 2 \varkappa k \left( \Upsilon_1 - \Upsilon_2 \right) +
  \left[
    - \omega^2 \mp 2 \varkappa \omega \left( \Upsilon_1 + \Upsilon_2 \right)
  \right] = 0.
\label{1-3-90}
\end{equation}

Here we have two cases:
\begin{enumerate}
  \item[(1)] For $A_2 = i A_1$, the wave vector is
  \begin{equation}
  k_{+,1,2} = - \varkappa \left( \Upsilon_1 - \Upsilon_2 \right) \pm
  \sqrt{
    \varkappa^2 \left( \Upsilon_1 - \Upsilon_2 \right)^2 +
    \omega^2 +
    2 \varkappa \omega \left( \Upsilon_1 + \Upsilon_2 \right)
  }.
  \label{1-3-100}
  \end{equation}
  \item[(2)] For $A_2 = - i A_1$, the wave vector is
  \begin{equation}
  k_{-,1,2} = \varkappa \left( \Upsilon_1 - \Upsilon_2 \right) \pm
  \sqrt{
    \varkappa^2 \left( \Upsilon_1 - \Upsilon_2 \right)^2 +
    \omega^2 -
    2 \varkappa \omega \left( \Upsilon_1 + \Upsilon_2 \right).
  }
\label{1-3-110}
\end{equation}
\end{enumerate}
Thus, we see that in both cases there are two GWs with different wave vectors for the same frequency $\omega$.
But in the second case a situation may occur where the GW is damped. This happens if
\begin{equation}
  \omega^2 - 2 \varkappa \omega \left( \Upsilon_1 + \Upsilon_2 \right) + \varkappa^2 \left( \Upsilon_1 - \Upsilon_2 \right)^2 < 0,
\label{1-3-120}
\end{equation}
and the GW becomes damped when $\omega$ lies in the region
\begin{equation}
  \varkappa \left(
    \sqrt \Upsilon_1 - \sqrt \Upsilon_2
  \right)^2 < \omega < \varkappa \left(
    \sqrt \Upsilon_1 + \sqrt \Upsilon_2
  \right)^2 .
\label{1-3-140}
\end{equation}
Let us consider the simplest case, when
$\Upsilon_1 = \Upsilon_2 = \Upsilon$. In this case
\begin{equation}
	k_{\pm, 1,2} = \pm \sqrt{\omega^2 \pm 4 \varkappa \omega \Upsilon},
\label{1-3-150}
\end{equation}
and for the sign $(-)$ a situation may occur where the GW does not exist.
For this case the phase and group velocities of the GWs will be the same as those in the case I from  subsection~\ref{case1}.

\section{Bianchi identities}

Now check the Bianchi identities for Eq.~\eqref{2-0-10},
\begin{equation}
  \left\langle
    \widehat{\delta T}_{\bar a \phantom{\mu}}^{\phantom{a} \mu}
  \right\rangle_{; \mu} = \frac{\partial
  \left\langle
    \delta \hat T_{\bar a}^{\phantom{a} \mu}
  \right\rangle }{\partial x^\mu} = 0.
\label{5-10}
\end{equation}
Here we took into account that the covariant derivative
$(\cdots)_{;\mu}$ is calculated in Minkowski spacetime. For the case I we have the following expression for
$\widehat{\delta T}_{\bar a \phantom{\mu}}^{\phantom{a} \mu}$:
\begin{equation}
 \widehat{\delta T}_{\bar a \phantom{\mu}}^{\phantom{a} \mu} =
 \begin{pmatrix}
		0	&	0	&	0	&	0 	\\
		0	&	0	&	0	&	0 	\\		
		0	&	0	&	
		- 4 \Upsilon_{1}\dot h_{\bar y \bar z}	&	
		4 \Upsilon_{1} \dot h_{\bar y \bar y} 	\\
		0	&	0	&	
		4 \Upsilon_{1} \dot h_{\bar y \bar y}	&	
		4 \Upsilon_{1} \dot h_{\bar y \bar z} 	
	\end{pmatrix}
\label{5-20}
\end{equation}
with $\Upsilon_{1}$ taken from Eqs.~\eqref{1a-3-50} and \eqref{1a-3-60}.

For the case II we have
\begin{equation}
 \widehat{\delta T}_{\bar a \phantom{\mu}}^{\phantom{a} \mu} =
 \begin{pmatrix}
		0	&	0	&	0	&	0 	\\
		0	&	0	&	0	&	0 	\\		
		0	&	0	&	
		- 2 \left[ \left(
  \Upsilon_1 - \Upsilon_2
 \right) h_{\bar y \bar z}^{\prime} + \left(
  \Upsilon_1 + \Upsilon_2
 \right)\dot{h}_{\bar y \bar z} \right] &	
		2 \left[ \left(
  \Upsilon_1 - \Upsilon_2
 \right)h_{\bar y \bar y}^{\prime} +
 \left(
  \Upsilon_1 + \Upsilon_2
 \right)\dot{h}_{\bar y \bar y} \right] \\
		0	&	0	&	
		2 \left[ \left(
  \Upsilon_1 - \Upsilon_2
 \right)h_{\bar y \bar y}^{\prime} +
 \left(
  \Upsilon_1 + \Upsilon_2
 \right)\dot{h}_{\bar y \bar y} \right]	&	
		2 \left[ \left(
  \Upsilon_1 - \Upsilon_2
 \right) h_{\bar y \bar z}^{\prime} + \left(
  \Upsilon_1 + \Upsilon_2
 \right)\dot{h}_{\bar y \bar z} \right] 	
	\end{pmatrix}
\label{5-30}
\end{equation}
with $\Upsilon_{1,2}$ given by Eqs.~\eqref{2b-12} and \eqref{2b-14}. For both cases one can show by direct calculation that
\begin{equation}
  \frac{\partial
  \left\langle
    \delta \hat T_{\bar a}^{\phantom{a} \mu}
  \right\rangle }{\partial x^\mu} = 0.
\label{5-40}
\end{equation}

It is interesting that in both cases operations of evaluating the covariant derivative and the quantum averaging commutate.
To show this,
let us calculate the averaged Bianchi identities
\begin{equation}
  \left\langle \left(
  		\widehat{\delta T}_{\bar a \phantom{\mu}}^{\phantom{a} \mu}
  \right)_{; \mu}
  \right\rangle = \left\langle
		\frac{ \partial
  		\widehat{\delta T}_{\bar a \phantom{\mu}}^{\phantom{a} \mu}
  		}{\partial x^\mu} +
  		\Gamma^\mu_{\nu \mu}
  		\widehat{\delta T}_{\bar a \phantom{\nu}}^{\phantom{a} \nu} -
  		\omega^{\bar c}_{\phantom c \bar a \nu}
  		\widehat{\delta T}_{\bar c \phantom{\nu}}^{\phantom{c} \nu}
  \right\rangle =
  \left\langle
		\frac{ \partial
  		\widehat{\delta T}_{\bar a \phantom{\mu}}^{\phantom{a} \mu}
  		}{\partial x^\mu}
  \right\rangle =
  \frac{\partial
  \left\langle
    \delta \hat T_{\bar a}^{\phantom{a} \mu}
  \right\rangle }{\partial x^\mu} = 0.
\label{5-50}
\end{equation}
Here we took into account that both the unperturbed Christoffel symbols
$\Gamma^\alpha_{\beta \gamma} = 0$ and the unperturbed spin connection
$\omega_{\bar a \bar b \mu} = 0$, since they are calculated for Minkowski spacetime.

\section{Conclusions}

We have considered the process of propagation of GWs on the background of the nonperturbative vacuum of spinor fields.
Using the simplifying assumptions from Sec.~\ref{RHS_En}, it was shown that there are
several distinctive features in comparison with the propagation of GWs through empty space:
\begin{itemize}
  \item There exists the fixed phase difference of $\pm \pi/2$
  between components $h_{yy, zz}$ and $h_{yz}$.
  \item The phase and group velocities of GWs are not equal to the velocity of light. Moreover, the group velocity is always less than
  the velocity of light.
  \item The components $h_{yy, zz}$ and $h_{yz}$ exist together only.
  \item Depending on the properties of the spinor vacuum,
  the damping of GWs may occur for some frequencies $\omega$
  of the spinor field, or no GW may exist.
  \item For given frequency $\omega$, there exist two waves with different wave vectors $k$.
\end{itemize}

All features mentioned above can in principle be verified
after the experimental detection of GWs. Then the simplest test
will be to verify the existence of the phase difference.
In addition, one might expect that GWs could be a fruitful tool for studying \emph{nonperturbative} quantum field theories.

\section*{Acknowledgments}

This work was partially supported  by grants \#1626/GF3 and \#139
in fundamental research in natural sciences by the Science Committee of the Ministry of Education and Science of Kazakhstan and by a grant of the Volkswagen Foundation.

\appendix

\section{Perturbed energy-momentum tensor}
\label{app1}

The perturbed component of the energy-momentum tensor is calculated as follows:
\begin{equation}
  \widehat{\delta T}_{\bar a \bar b} = - \frac{i}{2} \hat{\bar \psi} \left[
  		\gamma_{( \bar a} \delta \Gamma_{\bar b )} +
  		\delta \Gamma_{( \bar a} \gamma_{\bar b )}
  \right] \hat \psi,
\label{1-2-80}
\end{equation}
where the perturbed spinor connection is
\begin{equation}
  \delta \Gamma_{\bar a} = \delta \left(
  		e_{\bar a}^{\phantom a \mu} \Gamma_\mu
  \right) =
  - \frac{1}{4} \left(
  		\delta e_{\bar a}^{\phantom{a} \mu} \omega_{\bar b \bar c \mu} +
  		e_{\bar a}^{\phantom{a} \mu} \delta \omega_{\bar b \bar c \mu}
  \right) \gamma^{\bar b} \gamma^{\bar c},
\label{1-2-90}
\end{equation}
where the spin connection
$
\Gamma_\mu = - \frac{1}{4} \omega_{\bar a \bar b \mu}
\gamma^{\bar a} \gamma^{\bar b}
$. The perturbed vierbein $\delta e_{\bar a}^{\phantom{a} \mu}$ is
\begin{equation}
  \delta e_{\bar a}^{\phantom{a} \mu} =
	\begin{pmatrix}
		0	&	0	&	0	&	0 	\\
		0	&	0	&	0	&	0 	\\
		0	&	0	&	h_{\bar y \bar y}	&	h_{\bar y \bar z} 	\\
		0	&	0	&	h_{\bar z \bar y}	&	h_{\bar z \bar z} 	
	\end{pmatrix}	.
\label{1-2-82}
\end{equation}
Using the standard definitions of the covariant derivative of a spinor, $\nabla_\mu$, and the spin connection,
$\Gamma_\mu, \omega_{\bar a \bar b \mu}$,
\begin{eqnarray}
	\nabla_\mu \psi &=&
	\frac{\partial \psi}{\partial x^\mu} - \Gamma_\mu \psi,
\label{a2-10}\\
	\Gamma_\mu &=& - \frac{1}{4} \omega_{\bar a \bar b \mu}
	\gamma^{\bar a} \gamma^{\bar b},
\label{a2-15}\\
	\omega_{\bar a \bar b \mu} &=& - e_{\bar a}^{\phantom{a} \alpha}
	e_{\bar b}^{\phantom{b} \beta} \Delta_{\alpha \beta \mu}
\label{1-2-125}
\end{eqnarray}
with
\begin{eqnarray}
\label{a2-30}\\
	\Delta_{\alpha \beta \gamma} &=&
	e_{\bar a \alpha} \Sigma^{\bar a}_{\phantom{a} \beta \gamma} -
	e_{\bar a \beta} \Sigma^{\bar a}_{\phantom{a} \alpha \gamma} -
	e_{\bar a \gamma} \Sigma^{\bar a}_{\phantom{a} \alpha \beta} ,
\label{a2-40}\\
	\Sigma^{\bar a}_{\phantom{a} \alpha \beta} &=&
	\frac{1}{2} \left(
		\frac{\partial e^{\bar a}_{\phantom{a} \mu}}{\partial x^\nu} -
		\frac{\partial e^{\bar a}_{\phantom{a} \nu}}{\partial x^\mu}
	\right),
\label{a2-50}
\end{eqnarray}
one can obtain the perturbed spin connection
\begin{equation}
	\delta \omega_{\bar a \bar b \mu} =
	- \delta e_{\bar a}^{\phantom{a} \alpha}
	e_{\bar b}^{\phantom{b} \beta} \Delta_{\alpha \beta \mu} -
	e_{\bar a}^{\phantom{a} \alpha}
	\delta e_{\bar b}^{\phantom{b} \beta} \Delta_{\alpha \beta \mu} -
	e_{\bar a}^{\phantom{a} \alpha}
	e_{\bar b}^{\phantom{b} \beta} \delta \Delta_{\alpha \beta \mu}
\label{a2-65}
\end{equation}
with
\begin{eqnarray}
\label{a2-60}\\
	\delta \Delta_{\alpha \beta \gamma} &=&
	\delta e_{\bar a \alpha} \Sigma^{\bar a}_{\phantom{a} \beta \gamma} +
	e_{\bar a \alpha} \delta \Sigma^{\bar a}_{\phantom{a} \beta \gamma} -
	\delta e_{\bar a \beta} \Sigma^{\bar a}_{\phantom{a} \alpha \gamma} -
	e_{\bar a \beta} \delta \Sigma^{\bar a}_{\phantom{a} \alpha \gamma} -
	\delta e_{\bar a \gamma} \Sigma^{\bar a}_{\phantom{a} \alpha \beta} -
	e_{\bar a \gamma} \delta \Sigma^{\bar a}_{\phantom{a} \alpha \beta} ,
\label{a2-70}\\
	\delta \Sigma^{\bar a}_{\phantom{a} \alpha \beta} &=&
	\frac{1}{2} \left(
		\frac{\partial \delta e^{\bar a}_{\phantom{a} \mu}}{\partial x^\nu} -
		\frac{\partial \delta e^{\bar a}_{\phantom{a} \nu}}{\partial x^\mu}
	\right) .
\label{a2-80}
\end{eqnarray}
Here $\Delta_{\alpha \beta \gamma}$ and
$\delta \Delta_{\alpha \beta \gamma}$ are the unperturbed and perturbed Ricci coefficients; $\Sigma^{\bar a}_{\phantom{a} \alpha \beta}$ and
$\delta \Sigma^{\bar a}_{\phantom{a} \alpha \beta}$ are the unperturbed and perturbed anholonomy coefficients;
$\delta \omega_{\bar a \bar b \mu}$ are the perturbed spin connection.

Substituting \eqref{1-2-82} into \eqref{a2-65} and taking into account \eqref{a2-70} and \eqref{a2-80}, we have
\begin{eqnarray}
  \delta \omega_{\bar t \bar y y} &=&
  \delta \omega_{\bar x \bar y y} =
  - \delta \omega_{\bar t \bar z z} =
  - \delta \omega_{\bar x \bar z z} =
  \dot h_{\bar y \bar y},
\label{1-2-110}\\
  \delta \omega_{\bar t \bar z y} &=&
  \delta \omega_{\bar x \bar z y} =
  \delta \omega_{\bar t \bar y z} =
  \delta \omega_{\bar x \bar y z} =
  \dot h_{\bar y \bar z}.
\label{1-2-120}
\end{eqnarray}

\end{document}